\documentclass{ws-ijmpd}
\usepackage{graphicx}
\usepackage{makecell}
\usepackage{here}
\usepackage{float}
\usepackage{scalerel}

\numberwithin{equation}{section}

\usepackage[super,compress]{cite}
\begin{document}

\markboth{A. C. Khunt, V. O. Thomas and P. C. Vinodkumar}
{Distinct Classes of Compact Stars Based On Geometrically Deduced Equations of State}

%%%%%%%%%%%%%%%%%%%%% Publisher's Area please ignore %%%%%%%%%%%%%%%
%
\catchline{}{}{}{}{}
%
%%%%%%%%%%%%%%%%%%%%%%%%%%%%%%%%%%%%%%%%%%%%%%%%%%%%%%%%%%%%%%%%%%%%

\title{ DISTINCT CLASSES OF COMPACT STARS BASED ON GEOMETRICALLY DEDUCED EQUATIONS OF STATE  }

\author{A. C. Khunt}

\address{Department of Physics, Sardar Patel University,\\
Vallabh Vidyanagar-388 120, Gujarat,
 INDIA,
\\
ankitkhunt@spuvvn.edu}

\author{V. O. Thomas}

\address{Department of Mathematics, Faculty of Science,\\The Maharaja Sayajirao University of Baroda,
\\Vadodara – 390 001, Gujarat, India
\\
votmsu@gmail.com}

\author{P. C. Vinodkumar}

\address{Department of Physics, Sardar Patel University,\\
Vallabh Vidyanagar-388 120, Gujarat,
 INDIA,
\\
p.c.vinodkumar@gmail.com}
\maketitle

\begin{history}
\received{Day Month Year}
\revised{Day Month Year}
\end{history}

%%%%%%%%%%%%%%%%%%%%%%%%%%%%%%%%%%% ABSTRACT%%%%%%%%%%%%%%%%%%%%%%%%%%%%
\begin{abstract}
 We have computed the properties of  compact objects like neutron stars based on equation of state (EOS) deduced from a core-envelope model of superdense stars. Such superdense stars have been studied by solving the Einstein's equation based on pseudo-spheroidal and spherically symmetric space-time geometry. The computed star properties are compared  with those obtained based on nuclear matter equations of state. From the mass-radius ($M-R$) relationship obtained here, we are able to classify compact stars in three categories: (i) highly compact self -bound stars that represents exotic matter compositions with radius lying below 9 km (ii) normal neutron stars with radius between 9 to 12 km and (iii) soft matter neutron stars having radius lying between 12 to 20 km. Other properties such as Keplerian frequency, surface gravity and surface gravitational redshift are also computed for all the three types. The present work would be useful for the study of highly compact neutron like stars having exotic matter compositions.

\end{abstract}


\ccode{PACS numbers:}

%\tableofcontents
%%%%%%%%%%%%%%%%%%%% INTRODUCTION %%%%%%%%%%%%%%%%%%%%%%%%%%%%%%%%%%%%%%%%
\section{Introduction}	
Neutron stars are one of the densest objects in the observable universe. It represents state of  matter with highest densities. As such, they are valuable laboratories for  the study of dense matter. Such studies include  interplay between various disciplines like general relativity, high-energy astrophysics, nuclear and particle physics etc \cite{lattimer,pote}. Neutron stars have masses
of about (1$\sim $3 \(\textup{M}_\odot\)). These stars with masses about 1.2 \(\textup{M}_\odot\)  have central densities more than normal nuclear matter density and  radius of the order of 10 km. The average mass density $\rho$ of  the neutron star is approximately $10^{15}$ g cm$^{-3}$, which is about 3 times  the nuclear saturation density $\rho_{n}$$=2.7 \times 10$$^{14}$ g cm$^{-3}$ and at the core $\rho$ $>$ $\rho_{n}$ \cite{glend}. The magnetic field of such a compact stars lies  between $10^{8}$ -10 $^{15}$ G and possess gravitational field  $2\times 10^{11}$ cm s$^{-2}$ times stronger than that of earth's gravitational fields. The structure of these stars can be considered having an outer and  an inner crust. The envelope (outer crust) matter consists of  atomic nuclei (ions) and electrons. The thickness of envelope is few hundred meters. The inner crust occurs at a density of $4\times 10^{11}$g cm$^{-3}$ which consists of electrons, free neutrons and neutron-rich atomic nuclei. The thickness of this crust is typically about few kilometers. The outer crust envelopes the inner crust, which expands from the neutron drip density to a transition density $\rho$
$_{tr}$ $\sim$ $1.0 \times 10^{14}$ g cm$^{-3}$. And beyond the transition density one enters the \textit{core}, where all atomic nuclei have been melt down into their components, neutrons and protons. Caused by the high density and Fermi pressure, the core might also contain more massive baryon resonances or possibly a gas of free up, down and  strange quarks. Ultimately, $\pi$ and K mesons condensates may be found there too. All these dissimilar internal structure lead to different physical equation of state and hence contrasting mass-radius (M-R) relations. \par

In view of our inadequate knowledge of the equation of state of matter at extremely high densities, when matter density of ultra dense spherical objects is much higher than nuclear saturation density ($\rho$ $>$ $\rho_{n}$), it is difficult to have proper elucidation of matter in the form of an equation of state and quantitative calculations for the structure of neutron stars become obscure. A methodical valuation on the structure and properties of neutron stars can be found in \cite{pote,datta,heis,latti2,latti3,ilona,chamel,book}(and references therein). Many theorists have developed theoretical models for the structure of neutron stars which may be made up of various layers including core (inner and outer), crust (inner and outer) in which atomic nuclei are arranged into a crystal and the liquid ocean composed of the coulomb fluid \cite{pote}. The central region, i.e., core contains hyperons\cite{balb,weiss} or quark matter \cite{wittin}. A detailed analysis of quarks core models are discussed by Bordbar, Bigdell and Yazdizadeh \cite{bordbar}. Alternative method to study compact high-density astrophysical objects is through the space-time metric of the general theory of relativity and solving the relevant Einstein's equations. Such attempts particularly for compact object, like  the neutron star exist\cite{vaidya}. Thus, for the present study we make use of the core-envelope model for neutron stars studied based on the geometric approach making use of the relativistic model for these regions. The core-envelope model of a neutron star\cite{thomas,gedela,mafa} has different physical properties in envelope and core regions. From this we have considered two different EOS, based on anistropic pressure in core or envelope region. The core-envelope models studied by Thomas, Ratanpal and Vinodkumar\cite{thomas}(TRV model)
have considered anisotropic pressure in the envelope region and isotropic pressure in the core region. While in another case studied by S. Gedela, N. Pant, J. Upreti and R. Pant (SNJR model)\cite{gedela} have taken both  the core and envelope region as anistropic. In both the cases valid solutions of the Einstein's equations were studied in appropriate metrics. The EOSs deduced from these models are then used to compute the neutron star properties. Brief descriptions of these two models \cite{thomas, gedela} are given in the following section.

%%%%%%%%%%%%%%%%%%%%% core-envelope framework %%%%%%%%%%%

\section{Relativistic core-envelope framework}
 Our primary focus in this paper is based on the models belonging to the core-envelope family as discussed by Thomas et al. (TRV) and Gedela et al. (SNJR). We summarize below only the relevant part of the formalism  adopted for the study of compact objects with appropriate geometric consideration. More details can be found in the earlier works  \cite{thomas,gedela}
 
 \noindent  A nonrotating spherical metric in a most general form can be expressed as \cite{kip}
\begin{equation}
    ds^{2}= e^{\nu(r)}dt^{2}-e^{\lambda(r)}dr^{2}-r^{2}d\theta^{2}-r^{2}sin^{2}\theta d\phi^{2}
\end{equation}
 where, $r$ is the radial coordinate, $\theta$ is the polar angle and $\phi$ is the azimuthal angle. The right boundary condition for the metric is to match (2.1) with the Schwarchild exterior metric at the surface of the star. It is implemented as \cite{max}

\begin{equation}
    \nu(r=a)=\ln \bigg(1-\frac{2GM}{a c^{2}}\bigg)
\end{equation}

\begin{equation}
    \lambda(r=a)=-\ln \bigg(1-\frac{2GM}{a c^{2}}\bigg)
\end{equation}

 \noindent Here, $a$ and $M$ is the radius and mass of the star.

 \noindent The Einstein field equation is given by \cite{Steven}
\begin{equation}
    \mathfrak{R}_{\mu \nu}-\frac{1}{2} \mathfrak{R} g_{\mu \nu }=- \frac{8\pi G}{c^{4}} T_{\mu \nu}
\end{equation}

 \noindent has been solved for the metric given by Eqn (2.1)
for an energy momentum tensor relevant for perfect fluid \cite{thomas,gedela}

\begin{equation}
    T_{\mu \nu}=(\rho +p)u_{\mu}u_{\nu}-P g_{\mu \nu} +\pi_{\mu \nu}
\end{equation}
 where  $\pi_{\mu \nu}$ denotes anistropic stress tensor give by
 
 \begin{equation}
     \pi_{\mu \nu}=\sqrt{3} S \bigg[C_{\mu}C_{\nu}-\frac{1}{3}(u_{\mu}u_{\nu}-g_{\mu \nu})\bigg]
 \end{equation}
 where $S$=$S(r)$ is the magnitude of anisotropy stress tensor and $C^{\mu}=(0,  -e^{\frac{-\lambda}{2}},0,0)$, which is a radial vector.
 
  Thomas et al\cite{thomas}.have discussed core-envelope model on pseudo-spheroidal spacetime with core consisting of isotropic distribution of matter and envelope with anisotropic distribution of matter. While anisotropic core-envelope models by assuming linear equation of state in the core and quadratic equation of state in the envelope have been studied by Gedela et. al\cite{gedela}. In the following sub-sections we derive important aspects of these two models.
  \par

 %%%%%%%%%%%%%%%%% CORE-ENVELOPE %%%%%%%%%%%%%%%%%%%%%%%%%%%%%%%%%%%%%

\subsection{The TRV core-envelope model}
It has been shown that core and envelope regions consist of different physical features. They have chosen ansatz for a pseudo-spheroidal geometry of spacetime to solved the Einstein's equations. According to their metric, potential for pseudo-spheroidal geometry is expressed as

\begin{equation}
    e^{\lambda(r)}=\frac{1+K \frac{r^{2}}{R^{2}}}{1+\frac{r^{2}}{R^{2}}}
\end{equation}

 \noindent where $K$ and $R$ are geometric variables.

 \noindent The energy momentum tensor components (2.5) with anisotropic stress tensor $\pi_{\mu \nu}$ has non-vanishing components

\begin{equation}
    T^{0}_{0}=\rho, \,\,\,\,\ T^{1}_{1}=-\bigg(p+\frac{2S}{\sqrt{3}}\bigg),\,\,\,\,\ T^{2}_{2}=T^{3}_{3}=-\bigg(p-\frac{p}{\sqrt{3}}\bigg).
\end{equation}

  \noindent The magnitude of anistropic stress is give by \cite{thomas}
 \begin{equation}
     S=\frac{p_{r}-p_{\perp}}{\sqrt{3}}
 \end{equation}
 The boundary conditions for the core and envelope regions are

 \begin{equation}
   S(r)=0 \; \mbox{for} \;0\leq r\leq R_{C} \quad\text{and}\quad 
      S(r)\neq 0\;\; \mbox{for}\;  R_{C}\leq r\leq R_{E}
  \end{equation}
  
   \noindent where $R_{C}$ refers to the core boundary radius and $R_{E}$ corresponds to the envelope boundary radius which is the same as the radius of the star $(a)$ under consideration. Making use of these conditions with Eqn. (2.1), (2.4), (2.7) and (2.8), the Einstein field equations give the equations for density and pressure \\

  \noindent Accordingly, the density distribution ( core and envelope region ) is expressed as 
\begin{equation}
    \rho = \frac{1}{8\pi R^{2}}\bigg[ 3+2\frac{r^{2}}{R^{2}}\bigg] \bigg[1+2 \frac{r^{2}}{R^{2}}\bigg]^{-2}.
\end{equation}

 \noindent where R is a geometrical parameter. Equation (2.11) provides the density distribution in core and envelope region by using boundary condition for $0\leq r\leq R_{C}$ for core and $ R_{\scaleto{C}{3pt} }\leq r\leq R_{E}$ for envelope region.

\noindent The radial and transverse pressure in the envelope region is given by

\begin{equation}
    8\pi p_{\scaleto{E}{3pt}}= \frac{C\sqrt{1+\frac{r^{2}}{R^{2}}}\big(3+4\frac{r^{2}}{R^{2}}\big)+D} 
    {R^{2} \big(1+2\frac{r^{2}}{R^{2}}\big)^{2}\big(C \sqrt{1+\frac{r^{2}}{R^{2}}}+D\big)},
\end{equation}

\begin{equation}
    8\pi p_{\scaleto{E}{3pt}\bot}=   8\pi P_{\scaleto{E}{3pt}}
   -\frac{\frac{r^{2}}{R^{2}}\big(2-\frac{r^{2}}{R^{2}}\big)       }
  {R^{2} \big(1+2\frac{r^{2}}{R^{2}}\big)^{3}}.
\end{equation}
 and anisotropy $S$ has expression

 \begin{equation}
  8\pi\sqrt{3} S= \frac{ \frac{r^{2}}{R^{2}}\big(2-\frac{r^{2}}{R^{2}}\big)  }
  {R^{2}\big(1+2 \frac{r^{2}}{R^{2}}\big)^{3}}.
 \end{equation}

The constants $C$ and $D$ are given by

\begin{equation}
C=-\frac{1}{2}\bigg( 1+2\frac{a^{2}}{R^{2}}\bigg)^{-\frac{7}{4}},
\end{equation}

\begin{equation}
    D=\frac{1}{2}  \sqrt{1+\frac{a^{2}}{R^{2}}} \bigg(3+4 \frac{a^{2}}{R^{2}}\bigg) \bigg(1+2 \frac{a^{2}}{R^{2}}\bigg)^{-\frac{7}{4}}.
\end{equation}

\begin{figure}[H]
\centerline{\includegraphics[scale=0.50]{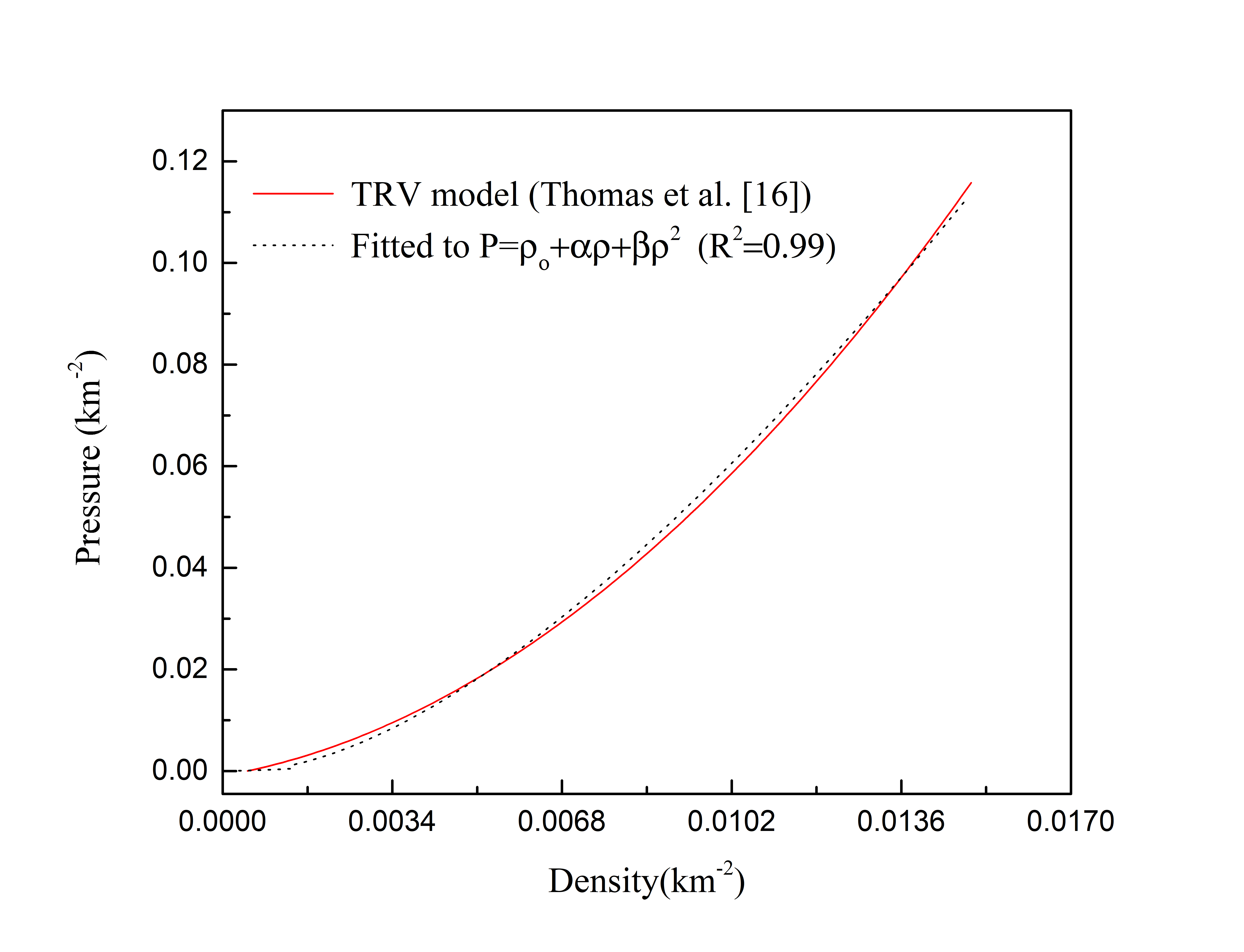}}
\vspace*{0pt}
\caption{(Color online) The radial pressure and density are given by Thomas et al.\cite{thomas}.(given in units of  km$^{-2}$), is plotted with solid curve. The dashed curve corresponds to the fitted curve with $\alpha= 406$, $\beta=1.69$ and $\rho_{0}=-9.30\times 10^{-4}$. For a density variation ($\lambda=0.01$).
 \label{f1}}
\end{figure}

\noindent The radial pressure in the core  region is given by

\begin{equation}
8\pi  p_{\scaleto{C}{3pt} }=  \frac{A\sqrt{1+\frac{r^{2}}{R^{2}}}+B \bigg[\sqrt{1+\frac{r^{2}}{R^{2}}}L(r)+\frac{1}{\sqrt{2}}\sqrt{1+2\frac{r^{2}}{R^{2}}}\bigg]}{ R^{2} \big(1+2\frac{r^{2}}  { R^{2}}\big) \bigg [A +\sqrt{1+2\frac{r^{2}}{R^{2}}}+B \bigg(\sqrt{1+\frac{r^{2}}{R^{2}}}  L(r) -\frac{1}{\sqrt{2}} \sqrt{1+ 2 \frac{r^{2}}{R^{2}}}\bigg) \bigg]}
\end{equation}
  where
\begin{equation*}
    L(r)= \ln\bigg(\sqrt{2}\sqrt{1+\frac{r^{2}}{R^{2}}} +\sqrt{1+2 \frac{r^{2}}{R^{2}}}\bigg).
\end{equation*}

where $A$ and $B$ are given by

\begin{equation}
    A=\frac{[5\sqrt{5}- 3\sqrt{2} (\sqrt{3} L(R_{c})-\sqrt 2.5)   ] C+ \frac{1}{\sqrt{3}} [5 \sqrt{5}+2\sqrt{2}(\sqrt{3} L(R_{c})-\sqrt{2.5}     ]      D       }
    {5\frac{5}{4}},
\end{equation}

\begin{equation}
    B= \frac{ \sqrt{2} } {5\frac{5}{4}} [3\sqrt{3}C-2D].
\end{equation}
 Equation (2.11) implies that the matter density at the center is explicitly related with geometrical variable $R$ as

\begin{alignat}{2}
 R=\sqrt{\frac{3\lambda}{8\pi \rho(a)}} &\quad , \:  \lambda=\frac{\rho(a)}{\rho(0)}= \frac{1+\frac{2 a^{2}}{3 R^{2}}}
  {(1+2\frac{a^{2}}{R^{2}})^{2}} 
\end{alignat}
We have plotted the graph of pressure against density in the TRV model and displayed by solid curve in  Fig.~\ref{f1} for density variation parameter $\lambda=0.01$. The best fit for the pressure-density curve is found to be in the quadratic from 
\begin{equation}
   p=\rho_{0}+\alpha\rho+\beta\rho^{2}
\end{equation}
where  $\rho_{0}=-9.30\times 10^{-4}$, $\alpha = 406 $  and $\beta=1.69$. It has been shown as a dotted curve in Figure 1. It can be shown that the model reveals  quadratic equation of state for different choices of the density variation parameter $\lambda$.

%%%%%%%%%%%%%%%%%%%  SNJR %%%%%%%%%%%%%%%%%%%%%%%%%%

\subsection{The SNJR core-envelope model}
In the second case of core-envelope anisotropic model of Gedela et al.\cite{gedela}, two distinct EOSs for core and envelope region are proposed.  For the core region $(0\leq r \leq R_{c})$, here a linear EOS as given below is used\cite{gedela}.

\begin{equation}
    p_{\scaleto{C}{3pt}}=(0.170)\rho-(7.833 \times 10^{-5})
\end{equation}

 \noindent The numerical values appeared in equation (2.22) are the same as given in\cite{gedela}. The expressions of density and pressure for core region are given by

\begin{equation}
    \rho_{\scaleto{C}{3pt}}= \frac{c(b r^{2}-3)}{8\pi (b r^{2}+1)^{3}}
\end{equation}

\begin{equation}
    p_{\scaleto{C}{3pt}}= \frac{c\alpha(b r^{2}-3)}{8\pi (br^{2}+1)^{3}}-\beta
\end{equation}

\noindent where $c$, $b$, $\alpha$ and $\beta$ are constants whose numerical values are $-0.00735$ km$^{-2}$, $0.0038$ km$^{-2}$, $0.1707$ km$^{-2}$ and $0.7833 \times 10^{-5}$ km$^{-2}$, respectively. For envelope region $(R_C\leq r \leq R_{E})$, they have considered quadratic EOS in the form

\begin{equation}
    p_{\scaleto{E}{3pt}}=\kappa\rho^{2}-\gamma
\end{equation}

\noindent where $\kappa$ and $\gamma$ are constants whose numerical values are $108$ km$^{-2}$ and $1.088 \times$ 10$^{-5}$ km$^{-2}$, respectively \cite{gedela}.

\noindent Further, the density and pressure profile in the envelope region are given by\cite{gedela} 
  
  \begin{equation}
      \rho_{\scaleto{E}{3pt}}= \frac{a(br-3)}{8\pi (br-1)^{3}}
  \end{equation}
\begin{equation}
    p_{\scaleto{E}{3pt}}=\frac{a^{2} \kappa(b r^{2}-3)^{3}}{64\pi^2 (br^{2}+1)^{6}}-\gamma
\end{equation}

\begin{figure}[H]
\centerline{\includegraphics[scale=0.50]{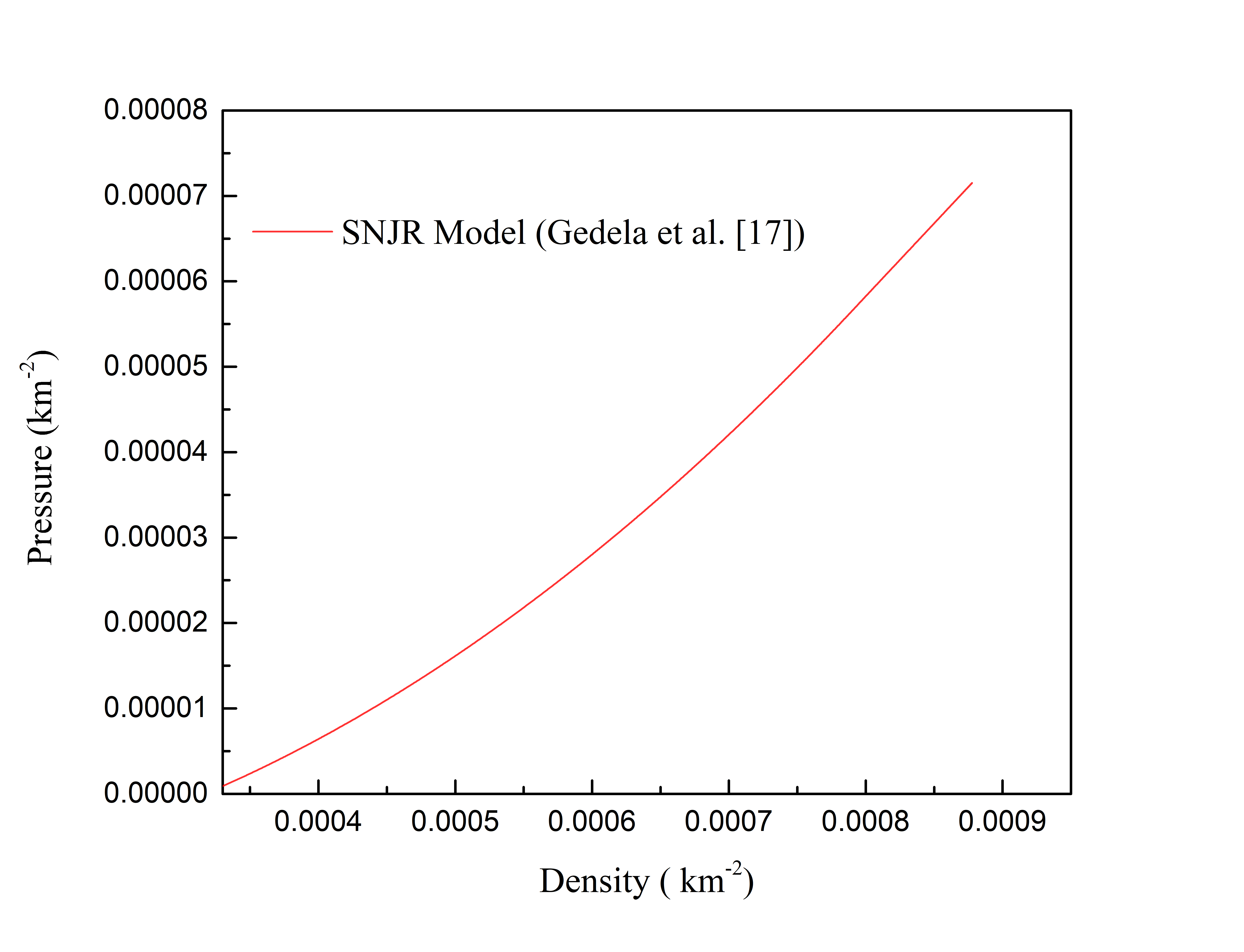}}
\vspace*{0pt}
\caption{ (Color online) Variation of a pressure $P$ in  (km$^{-2}$) with respect to a density $\rho$ in (km$^{-2}$) . Figure based on Eqn. (2.22) and (2.25) with $\kappa=$$108$ km$^{-2}$ and $\gamma =$$1.088 \times$ 10$^{-5}$ km$^{-2}$.
 \label{f2}}
\end{figure}
 \noindent An important feature of both of these core-envelope models (TRV and SNJR) is that they have the stable equilibrium under hydrostatic configuration. Theoretical study of the relativistic core-envelope model using  paraboloidal spacetime by Ratanpal and Sharma \cite{rat} have shown that paraboloidal geometry also admit quadratic equation of state. Other EOSs that we have considered in the present work for comparison include those considering different physical compositions of nuclear matter reported by \cite{alf, akmal, potekhin, eng, sly,sqm, wwf}. The different models used in this study are listed in Table~\ref{ta1}.\\
  \noindent  A very crucial feature of the equation of state is the causal limit( a sound signal cannot propagate faster than the speed of light, $\nu^{2}_{s}=dp/d\rho\leq c^{2}$). In both cases based on the geometrical models (TRV and SNJR) the causality condition is satisfied. In particular, Thomas et al.\cite{thomas} have studied the causality limit for different density variables. The computed  speed of sound ($\nu_{s}$) versus radius as shown in Figure 3. Both the cases clearly indicate the validity of causality condition.

  \begin{figure}[H]
\centerline{\includegraphics[scale=0.5]{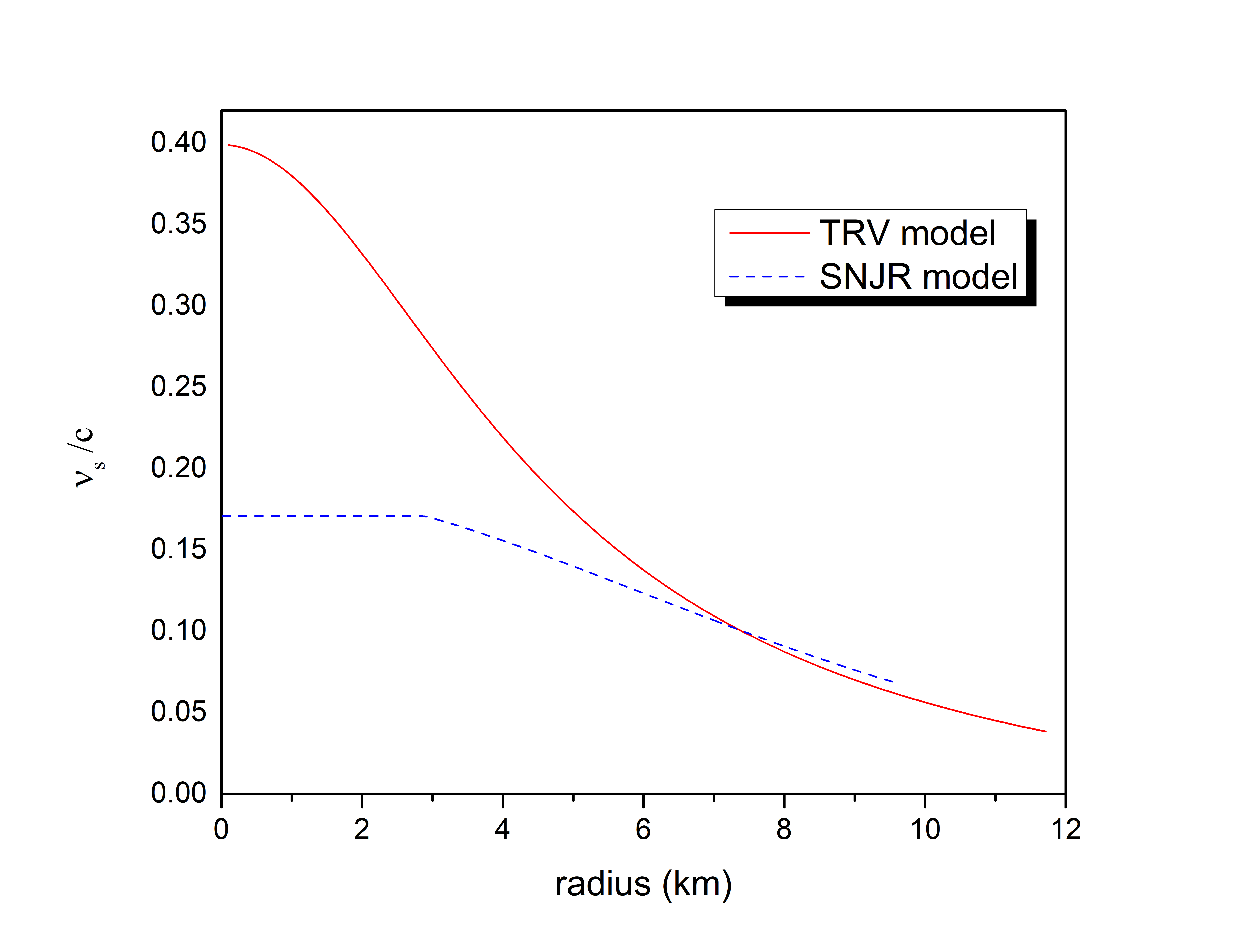}}
\vspace*{0pt}
\caption{(Color online) Velocity of sound, $\nu_{s}$, in unit of the speed of light, c, as a function of radius calculated for the TRV (Red solid line) equation of state (for $\lambda=0.01$) and SNJR (Blue dashed line) equation of state.
 }
\end{figure}

%%%%%%%%%%%% compact Star Structure %%%%%%%%%%%%%

\section{Compact Star Structure : Static Equilibrium configurations}

It is vital to explore static and spherical symmetrical gravity sources in general relativity, especially when it comes to internal structure of compact objects. For simplicity , we consider only nonrotating, spherically symmetric stars. The geometry inside the star is described by the familiar Tolman–Oppenheimer–Volkoff (TOV) equation, which is valid for a perfect fluid\cite{tov}. The equation of state is all that is required to solve the TOV equations. For static, spherically symmetric stars in hydrostatic equilibrium,
the TOV equations may be written as a pair of first-order differential equations. The calculation of neutron star structure is obtained by
numerically integrating the Tolman-Oppenheimer-Volkoff equation \cite{tov}
\begin{equation}
\frac{dP}{dr}=- \frac{G m(r)\rho(r)}{r^{2}}\frac{\big(1+\frac{P(r)}{\rho(r) c^{2}}\big)\big[1+\frac{4\pi r^{2}P(r)}{m(r)c^{2}}\big]}{1-\frac{2Gm(r)}{r c^{2}}},
\label{eqn1}
\end{equation}

\begin{equation}
\frac{d m(r)}{dr}=4\pi r^{2}\rho(r).    
\end{equation}
Here $P$ is the radial pressure, $\rho$ is the mass density, $r$ is the radial distance measured from the center, and $m(r)$ is the enclosed mass from
the center $r = 0$ where $P=P_{c}$ and $\rho=\rho_{c}$ to a radial distance $r$.  In the present work we have fixed the central density for both geometrical models at $\rho_{c}= 1.34\times 10^{15}$g cm$^{-3}$. The seven nuclear EOSs with a fixed central density at $\rho_{c}= 1.0\times 10^{15}$g cm$^{-3}$,
equations (3.1) and (3.2) are  integrated numerically to determine the global structure (e.g. radius and mass) of a neutron star.
\noindent To begin with, the density close to the center of the compact star is assumed to be homogeneous, with the density $\rho =\rho_{c}$, the radius $r=0.1$ cm and m(0.1 cm)= $4\pi \rho_{c} r^{3} /3 $. Equations (3.1) and (3.2) are integrated numerically from $r=0.1$ cm to the boundary of the star, where the pressure falls to zero  ($P(a)=0$ ). The total mass of the star is then given by  $M= m(a)$.\\
  \noindent Using the data files provided by Özel et al.\cite{ozel}, we have re-ploted pressure-density profile  corresponds to all the model EOS's  listed in Table~\ref{ta1} along with geometric EOS's of TRV and SNJR. It can be seen that  these EOS's distinctly differ from each other.

\begin{figure}[H]
\centerline{\includegraphics[scale=0.5]{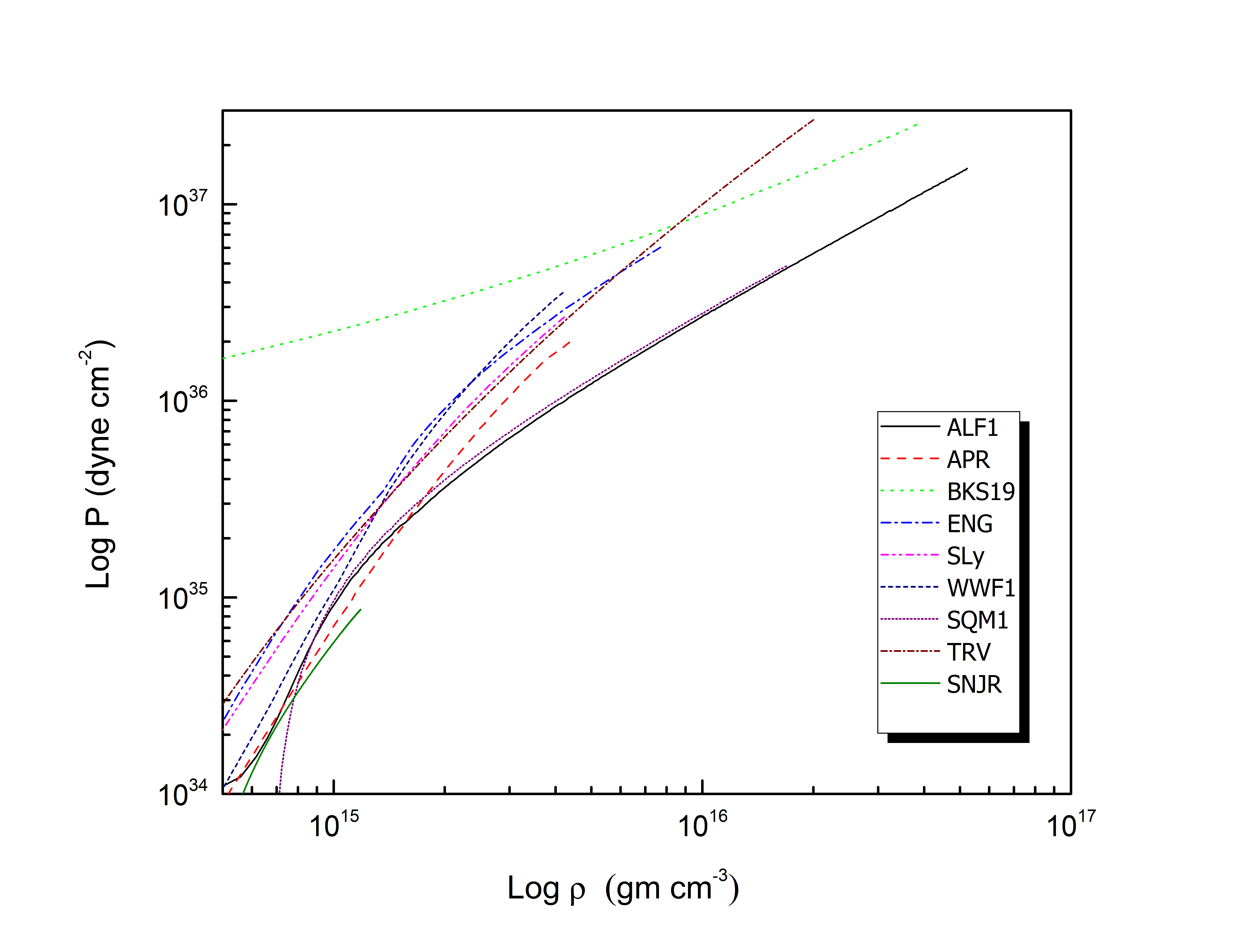}}
\vspace*{1pt}
\caption{(Color online) Geometrical EOS TRV( wine short dash line) and SNJR (olive solid line) compared to the selected nuclear EOS's ( ALF1 (black solid line), APR (red dash line), BKS19 (green dot line), ENG (blue dash dot line), SLy (magenta dash dot dot line), WWF1 (navy short dash line) and SQM1 (purple dot line). Details of these EOSs are listed in Table~\ref{ta1}.
 \label{f4}}
\end{figure}

%%%%%%%%%%%%%%%%%%%%%% M R %%%%%%%%%%%%%%%%%%%%%%%%%%%%%%%

\subsection{Mass-Radius Relation}
In this work, we have considered two general relativity inspired equations of state and  compared with seven different nuclear equations of states as listed in Table~\ref{ta1}. 
\\

\begin{table}[H]
\tbl{Nuclear and Geometrical equations of state used  for the construction of models of general relativistic static neutron stars}
{\begin{tabular}{@{}cccc@{}} \toprule
Label & EOS &  Composition and model &
 Reference \\

 \colrule
1 & ALF1 & nuclear plus quark matter (MIT Bag Model)  &  Alford et al. (2005)\cite{alf} \\  
& & & \\
2 & APR &  \makecell{$n p e \mu$, variational theory, Nijmegen NN \\ plus Urbana NNN potential}  & Akmal et al. (1998)\cite{akmal} \\
& & & \\
3 &\hphantom{0} BKS19 & \makecell{cold catalyzed nuclear matter \\analytical  unified EOSs} & Potekhin et al. (2013)\cite{potekhin}\\
& & & \\
4 & ENG &\makecell{Dirac-Brueckner HF\\asymmetric nuclear matter }  & Engvik et al. (1996)\cite{eng}\\ 
& & & \\

5 & SLy  & \makecell{potential method,  n p e $\mu $ \\effective nucleon energy functional  }  & \makecell{Douchin and \\ Haensel et al. (2001)\cite{sly}} \\ 
& & & \\

6& WWF1 & \makecell{variational method\\ dense nucleon matter}  &  Wiringa et al. (1988)\cite{wwf}\\ 

7 & SQM1 & \makecell{MIT Bag Model \\ (Strange quark matter)} & Zdunik  (2000)\cite{sqm}\\ 
\colrule
8 & TRV &\makecell{core   :   isotropic fluid distribution \\envelope : anistropic fluid distribution } & Thomas et al. (2005)\cite{thomas}\\ 
& & &\\
9 & SNJR & \makecell{core  :  linear equation of state \\envelope : quadratic equation of state} & Gedela et al. (2019)\cite{gedela}\\ 
\botrule
\end{tabular} \label{ta1}}
\end{table}

\noindent The composition and model used for all these equation of state and their respective bibliographic references are also listed in Table~\ref{ta1}.   Making use of these  equations of state, we obtained  the mass-radius relationship for a compact star. 

\noindent The mass-radius relations obtained with the help of nuclear equations of state of different compositions are compared with the geometrical equations of state and are plotted in Fig.~\ref{f4}. These plots reiterate the fact that  nuclear and geometrical equations of state  manifest three distinct types of compact stars. The first one corresponds to the two  cases represented by the models 7 and 8 of Table~\ref{ta1} , the second one corresponds to the models (1 to 6) largely represented by the nuclear matter EOSs and third type corresponds to the model (9) represented by the geometric model (SNJR). In all the three cases the maximum masses correspond to stable structure varies from 1.4 to 2.3 \(\textup{M}_\odot\), while the radius at their maximum masses lie 8 - 9 kms in the case of the first category, 9 - 12 kms in the cases of (second category) and  beyond 12 km in the case of the third category. The central density at maximum mass obtained here for the stable configurations are listed in Table~\ref{ta2}.

\begin{figure}[H]
\centerline{\includegraphics[scale=0.5]{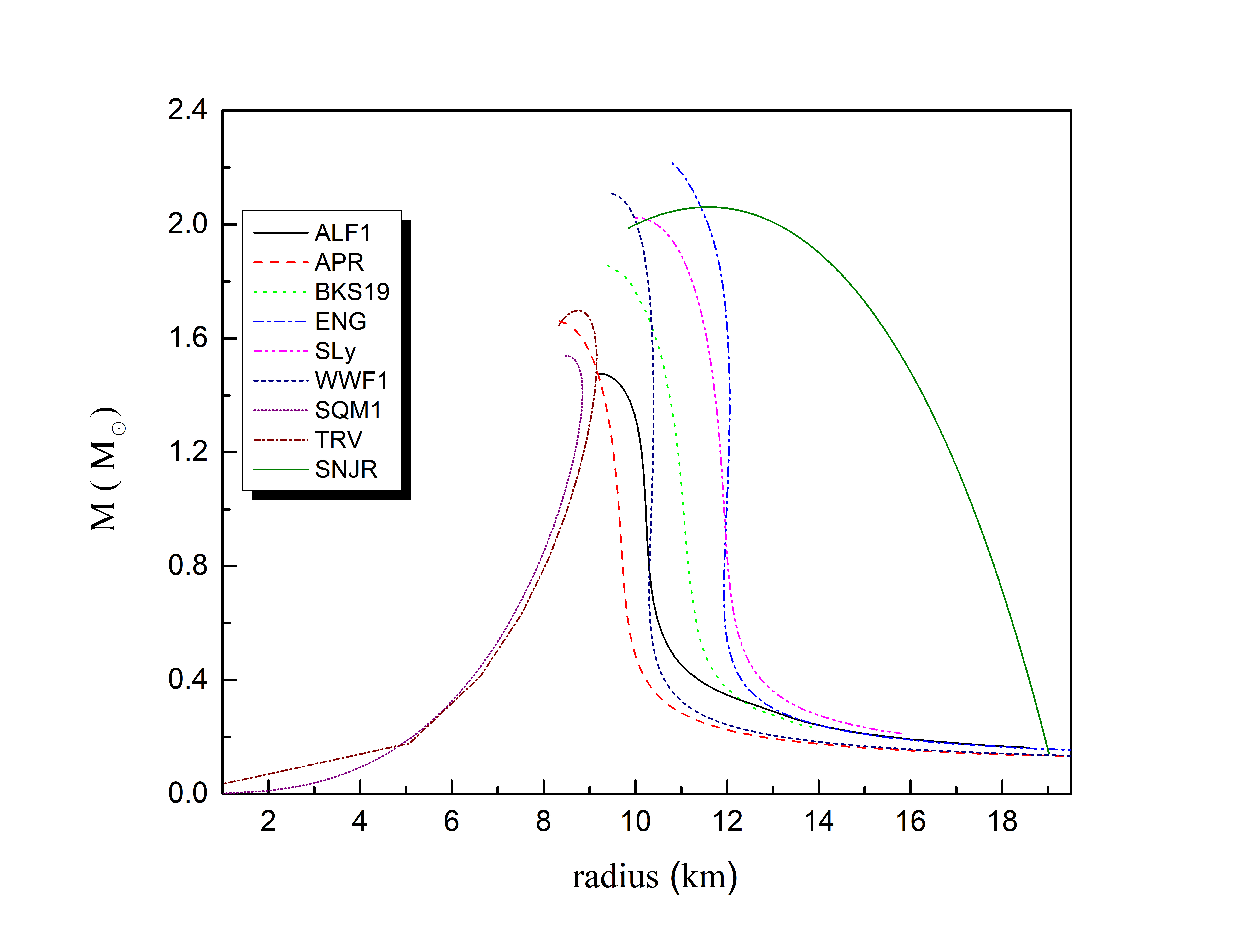}}
\vspace*{1pt}
\caption{ (Color online) Neutron star mass as a function of radii for pure nuclear matter EOSs vs. geometrical EOSs. The labels are explained in Table~\ref{ta1}.
 \label{f5}}
\end{figure}

\noindent The M-R diagram obtained from the two geometrically deduced models behave differently. We found that TRV equation of state resulted into the mass-radius curve similar to the one obtained for strange quark matter stars (SQM1, label-7)\cite{li}. The monotonically increasing mass with radius ($M \propto a^{3}$) is expected for the class of ultra compact objects which are self-bound\cite{milva}. The surface density of strange star is roughly fourteen orders of magnitude larger than the surface density of normal neutron stars \cite{weber}. The TRV model gives a stable configuration in the same orders of magnitude, with the surface density,  $\rho_{s}$ $\approx$ $2 \times 10^{14}$ g cm$^{-3}$. Thus, it is an appropriate geometrical model for the study of  ultra compact stars having exotic matter composition.

\noindent The isotropic fluid distribution in the core part of the TRV model\cite{thomas} is justified if the core matter distribution is of quarks or strange matter, governed by MIT bag model. Further, the envelope with anisotropic fluid distribution can be viewed as due to hadronization to baryonic matter. Thus the TRV model prediction fit well  with that of the strange quark matter stars with its maximum mass, ($M_{max}=1.69 $\(\textup{M}_\odot\)) and radius, 8.76 km. The SNJR model that predicts the third category in which EOS has  the linear behaviour inside the core and quadratic behaviour at the envelope has resulted into the M-R diagram  different from all other cases. Its M-R curve is broader as compared to all other cases studied here. And its density is much lower than that of normal neutron like stars. Recent observations of binary neutron-star mergers (GW170817) have reported an estimation for the radius of the  neutron star in the range 10.6 to 11.5 kilometers.\cite{collin}. 

,

\subsection{Keplerian frequency (rotation frequency of neutron star) }

 The Kepler frequency expresses the balance of centrifugal and gravitational force on a particle on equatorial plane at the surface of a star. It is expressed as

\begin{equation}
    \Omega_{c}=\sqrt{\frac{M}{a^{3}}} \; ,
\end{equation}
where the subscript $c$ denotes classical symmetry of the centrifugal and gravitational forces, which is the Newtonian expression for the Kepler angular velocity. This equation do to not hold in General Relativity, but as it turn out, it holds to very good accuracy if the right side is multiplied by a prefactor$(C)$ \cite{haensel}. It has been shown by J. M. Lattimer, et al\cite{lp}., Haensel et al.\cite{haensel} and B. Haskell et al.\cite{haskell} that the numerical value of the Keplerian frequency, namely the maximum rotational frequency of a neutron star accounting for the effects of general relativity, deformation , and independent on the EOS, can be well fitted from the simple formula 

\begin{equation}
      \Omega_{K} \approx C \bigg(\frac{M}{\text{\(M_\odot\)
}}\bigg)^{1/2}\bigg(\frac{10 \: \text{km}}{a}\bigg)^{3/2} \;  \text{Hz}\: ,
\end{equation}
 providing the neutron star mass is not very close to the maximum stable value, $M$ and $a$ are the mass and the radius of the nonrotating star respectively. The constant $C$ of Eq.3.4 are given by  B. Haskell et al.\cite{haskell}. For the self bound compact stars it is given as $1.15$ KHz and for other gravitationally bound neutron stars it is given as  $1.08$ KHz.
 
 \noindent The deduction of $\Omega_{K}$ generally requires the calculation of rotating general relativistic configurations. Nevertheless , Haensel et al. (2009) have shown  to a good degree of accuracy that the mass-shedding frequency $\Omega_{K, max}$ can be determined by the EOS-independent empirical formula  as given in Eq.(3.4). On the other hand, it allows to determine $\Omega_{K}$ using the mass and radius  of the nonrotating star.

\begin{figure}[ht!]
\centerline{\includegraphics[scale=0.5]{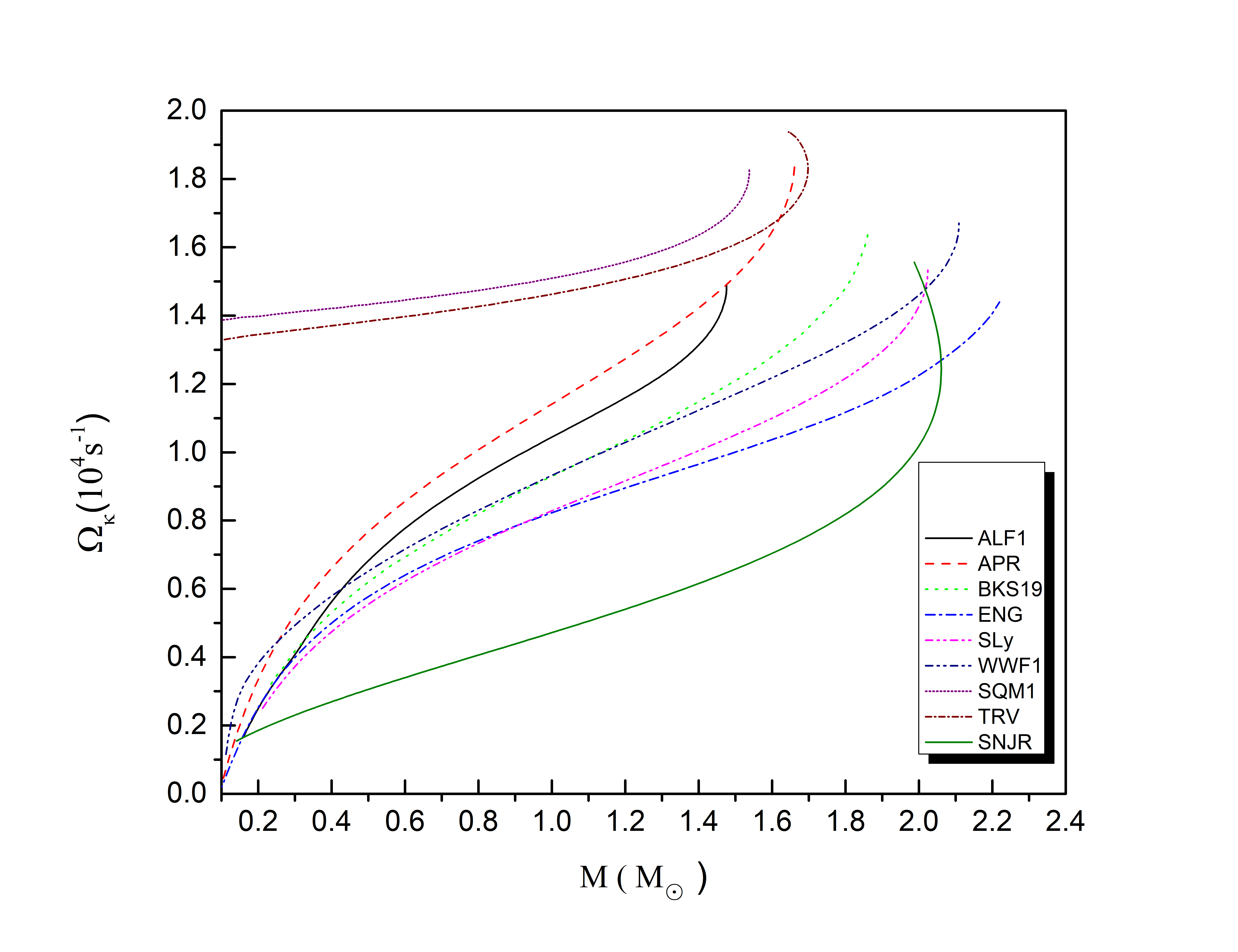}}
\vspace*{3pt}
\caption{ (Color online) Kepler frequency, $\Omega_{k}$, as a function of neutron star mass using the two different classes of EOS ( \textit{nuclear and geometrical})
 \label{f6}}
\end{figure}

\noindent The calculated Keplerian frequency based on the mass-radius relations obtained using all the nine equations of state  are shown in   Fig.~\ref{f6}.  Here we found that Keplerian frequency
corresponds to TRV and SQM1 are similar with higher values of $\Omega_{K}$ ( 14-18 KHz). While other cases $\Omega_{K}$ varies from 2 KHz to 18 KHz. The results of Keplerian frequency for the maximum mass of stable stars are shown in Table~\ref{ta2} for all the nine models.

\subsection{ Surface Gravity}
The surface gravity of neutron stars denoted by $g_{s}$ (i.e., the acceleration due to gravity as measured on the surface),  is an  important parameter for the study of neutron star atmospheres \cite{gud}. The upper bound of the surface gravity for neutron  stars with various baryonic EOSs is studied by Bejger et al. (2004)\cite{bejger}. The surface gravity of neutron star is  many orders of magnitude larger than that of other stars; it is $\sim$ 10$^{12}$ times stronger than gravity at the Earth surface, and $10^{5}$ times larger than that for the white dwarfs.\\
\\
The expression for $g_{s}$ is given by \cite{bejger} :

\begin{equation}
    g_{s}= \frac{G M}{a^2 \sqrt{1-x_{GR}}} 
\end{equation}

\noindent Here, $x_{GR}  =2 G M/a c^{2}=r_{g}/a$, where $r_{g}$ is the Schwarzschild radius.
The importance of relativistic effects for a neutron star mass $M$ and radius $a$ is characterized by the {\it compactness parameter} $r_{g}/a$.  Usually for a neutron star with $M=1.4$ \(\textup{M}_\odot\) and Radius is about 10 km, surface gravity becomes ($g_{s}$) = 2.43 $\times$ 10$^{14}$ cm s$^{-2}$. In consequence it is suitable to measure $g_{s}$ in units of 10$^{14}$ cm s$^{-2}$ and is represented as $g_{s, 14}$ $\equiv$  $g_{s}/(10^{14}$ cm s$^{-2}$). The computed surface gravity, $g_{s, 14}$ for all the cases studied here are shown in  Fig.~\ref{f8} against mass expressed in \(\textup{M}_\odot\). The numerically values of $g_{s, 14}$ correspond to maximum stable mass of the star are also listed in Table~\ref{ta2}.

\begin{figure}[h!]
\centerline{\includegraphics[scale=0.5]{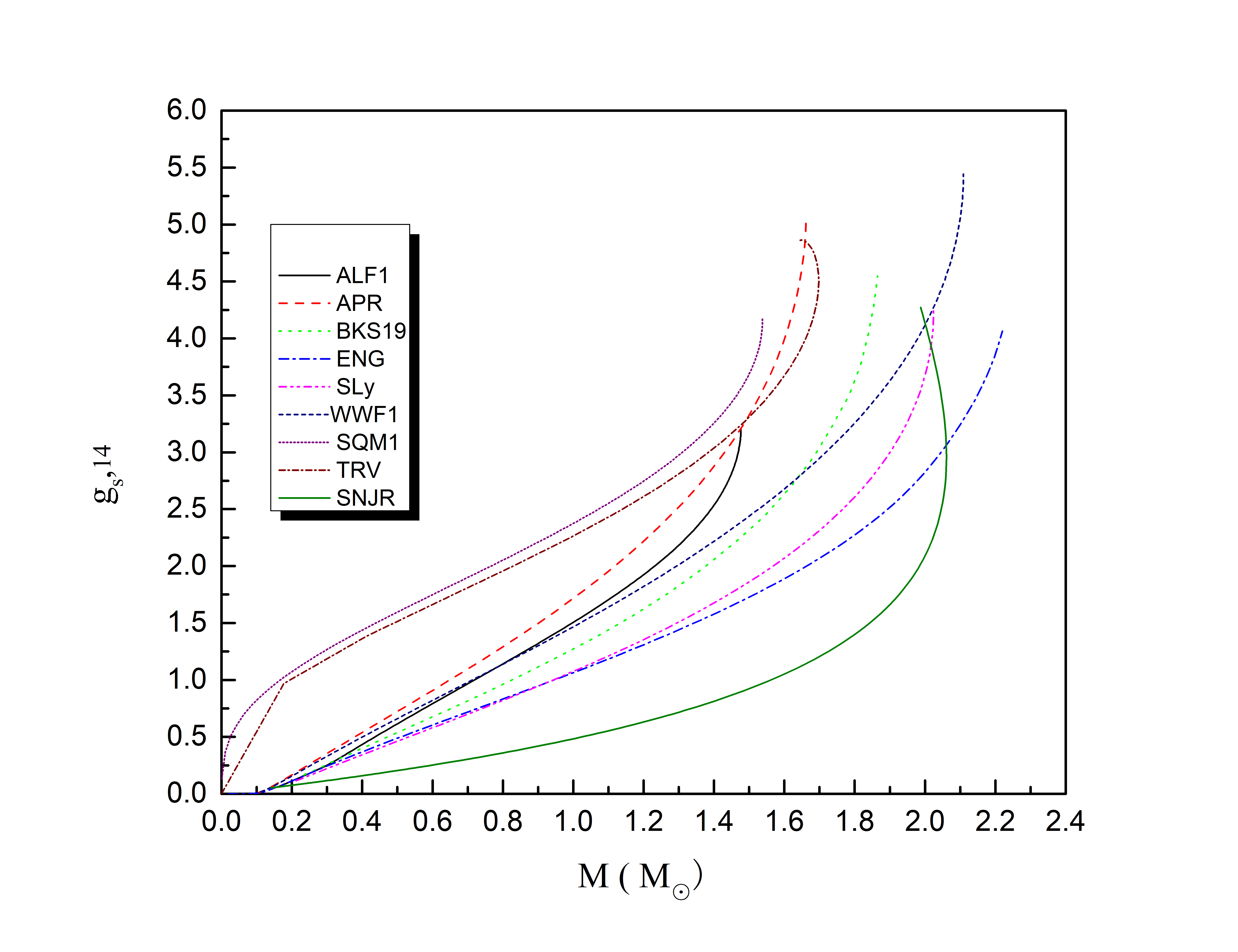}}
\vspace*{4pt}
\caption{(Color online) Plots of $g_{s, 14}$   versus gravitational mass $M$. Surface gravity in the units of 10$^{14}$ cm s$^{-2}$. \label{f7}}

\end{figure}

\noindent  It is found that for $M=1.4$ \(\textup{M}_\odot\), $g_{s, 14}$ ranges from 1.43 to 2.8 and for $M\approx 2.0$ \(\textup{M}_\odot\) the surface gravity lies between 1.88 to 4.38 . The nuclear EOSs (labeled : 1 and 6) with an exotic quarks phase have relatively low $g_{s, max}$. A similar situation occurs for the SNJR EOS that gives lowest value of surface gravity. The only reason SNJR EOS have low surface gravity is that they have a greater radius compared to other EOSs. The TRV EOS (labeled 8 ) yields $g_{s, max}$ similar to BKS19 and SLy EOSs. Their values of $g_{s,14 (max)}$ range from 4.10 to 4.60.  

\subsection{ Gravitational Redshift of Neuron Star}
In general relativity the ratio of the emitted wavelength $\lambda_{e}$ at the surface of a nonrotating neutron star to the observed wavelength $\lambda_{0}$ received at radial coordinate $r$, is given by $\lambda_{e}/\lambda_{0}=[g_{tt}(a)/g_{tt}(r)]^{1/2}$.  From this the definition of gravitational redshift, $z\equiv (\lambda_{0}-\lambda{e})/ \lambda{e}$ from the surface of
the neutron star as measured by a distant observer $( {g_{tt}(r)}\to -1)$ is given by 
\begin{equation}
    z=\mid - g_{tt}(a)\mid ^{-1/2}-1 =\bigg(1-\frac{2 G M}{a c^{2}}\bigg)^{-1/2}-1
\end{equation}
We compute the limit of the redshift from  the surface of a neutron star using Eqn (3.8) where $g_{tt}=-e^{\lambda(r)}=-(1-2 G M / c^{2}a)$ is the metric components \cite{tov}. For a given EOSs the  maximum value $z_{\text{surf}}^{\text {max}}$ increase with increase of $M_{\text{max}}$.  Neutron stars of $M\geq $ \(M_\odot\) are expected to have sizable $z_\text{surf}$. The computed values of $z_\text{surf}$ for all the cases studied here  are listed in Table~\ref{ta2}. The computed values of $z_\text{surf}$ are found to lie between $0.2$ to $0.3$.

\begin{figure}[H]
\centerline{\includegraphics[scale=0.4]{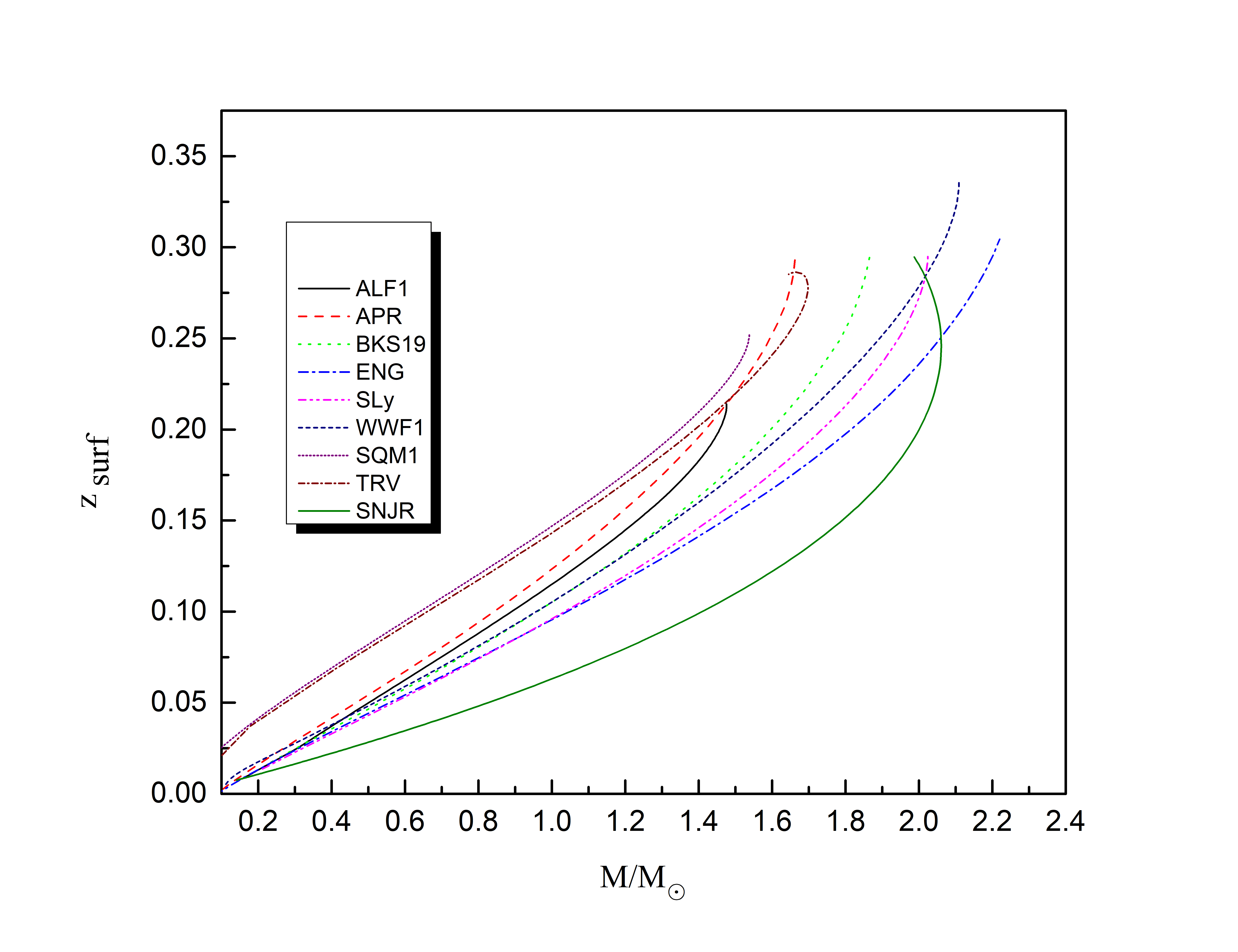}}
\vspace*{4pt}
\caption{ (Color online) Gravitational redshift at the neutron star surface as a function of the stellar gravitational mass for the nine considered EOS models. 
 \label{f8}}
\end{figure}

\noindent From the listed values of $x_{GR}$ in Table~\ref{ta2}, we found that all the models studied here satisfy the Buchdahl inequality\cite{buch}, $a\geq(9/8)r_{g} = (9/4)G M/c^{2}$   which is stricter than the Schwarzschild bound. A consequence of this is that the gravitational redshift should satisfy $z\leq 2$. The precise upper bound on the surface redshift for neutron star is $z_{\text{surf}}=0.851$ for subluminal EOSs \cite{gondek}. In the present study, we found that $z_\text{surf}$ for all the nine cases computed here lie much below the upper bound for the gravitational redshift.

\section{Results and Discussion }

We have computed several properties of a compact star like neutron star, using nuclear and geometrically deduced equations of state. We have used geometrical equations of state from the core envelope model, that describes different properties of the physics in the core and envelope region.  Many similarities and dissimilarities are observed from the properties computed based on the geometrical EOSs and the nuclear EOSs.

\begin{table}[H]
\tbl{Calculated properties of nonrotating neutron star models}
{\begin{tabular}{@{}cccccccc@{}} \toprule
Label & \makecell{$ M _{max}$ \\(\(M_\odot\))}  & \makecell{$a_{max}$ \\(km)} & \makecell{$\rho_{c}$\\($10^{15}$ gm cm$^{-3}$)}& \makecell{$\Omega_{k}$ \\ ($10^{4}$ s$^{-2}$) }
& $z_{\text{surf}}$   & \makecell{$g_{s, 14}$\\(cm s$^{-2}$)} & $x_{GR}$ \\

 \colrule
1 & 1.47 &  9.21& 3.34& 1.49 & 0.21 & 3.17& 0.47\\

2 & 1.65 & 8.37 &4.51& 1.85 & 0.28 & 4.89 &0.58 \\  

3 & 1.86 &  9.25 &2.69& 1.66 & 0.29 & 4.59 &0.59\\  

4 & 2.22 & 10.76  &2.91& 1.44 & 0.30 & 4.07 &0.60\\  

5& 2.00 & 10.08  &2.85& 1.53 & 0.29 & 4.14 &0.58\\  

6 & 1.73 &  9.17 &2.93& 1.67 & 0.26 & 5.36 & 0.65 \\ 

7 & 1.54 &  8.48 &3.27& 1.82 & 0.25 & 4.17 &0.53 \\

 \colrule
8 & 1.68 &  8.76 &3.58& 1.82 & 0.24 & 4.26 & 0.57 \\  

9 & 2.06 &  11.58 &0.61&1.24 & 0.23 & 1.98 & 0.56\\  

\botrule
\end{tabular} \label{ta2}}
\end{table}

\noindent In Table~\ref{ta2} we have listed computed properties of such a compact star with all the different types of EOSs. Like, maximum mass ($M_{max}$), stellar radius ($a_{max}$) correspond to the maximum mass , central density ($\rho_{c}$), Keplerian frequency ($\Omega_{k}$) correspond to the maximum mass for the stable structure of the star, gravitational redshift ($z_{\text{surf}}$), surface gravity ($g_{s}$) and compactness parameter ($x_{GR}$). We compared all these properties with the properties obtained from geometrically deduced equations states. The properties obtained from TRV equation of state are in good accordance with the properties obtained from other nuclear matter based models. While the parameters obtained using the SNJR model are quite different from others except for $z_\text{surf}$ and ($x_{GR}$). The central density that yields the maximum mass of $\approx$ 2 \(\textup{M}_\odot\)  in the case of SNJR is very low and the radius is about 12 km. It is also reflected in the low values of the surface gravity. It is noticed  that mass-radius configuration as shown in Fig.~\ref{f5} obtained  from geometrical models will be pertinent for divergent class of compact stars. Particularly,  the pseudo-spheroidal spacetime of TRV model seemed to describe the ultra dense compact stars like the strange self-bound stars. The spacetime geometry adopted for the SNJR model represents low density neutron  like star\cite{gedela}, where radius lie between $12\leq R \leq 20$ kms.

\noindent The mass-radius diagram  in  Fig.~\ref{f5} clearly classify the nature of compact stars in three categories :(i) highly compact self-bound stars represented by the TRV Model and SQM1 model with exotic matter compositions (ii) the normal neutron stars with nuclear matter EOS and (iii) the ultra soft compact stars represented by the SNJR geometrical model. At the end,  we are able to identify a correspondence between the geometric description with the structure of the matter distribution in compact objects like a strange star. To summarize, we have been able to classify neutron like compact stars in three  distinct types each one having different internal structures. We hope that TRV model for compact neutron like stars will be useful for the study of superdense self-bound stars having exotic matter compositions.

\section{Acknowledgements }
We would like to thank Feryal Özel  for kindly providing us with EOS table for the nuclear matter. Further , we thank Dr. B. S. Ratanpal for many helpful discussions.

\end{document}